\documentclass[a4paper,12pt]{article}
\usepackage{amsfonts}
\usepackage{latexsym}
\usepackage{amsmath}
\usepackage{amssymb}
\usepackage{amssymb}
\usepackage{slashed}
\usepackage{upgreek}
\usepackage{mathrsfs}
\usepackage{bbm}
\usepackage[symbol]{footmisc}

\newcommand*{\defeq}{\mathrel{\vcenter{\baselineskip0.5ex \lineskiplimit0pt
                     \hbox{\scriptsize.}\hbox{\scriptsize.}}}%
                     =}

\allowdisplaybreaks

\usepackage{relsize}
\usepackage{graphicx}
\usepackage{comment}

\usepackage{chngcntr}
\counterwithout{equation}{section}

\def\appendix#1{\addtocounter{section}{1}\setcounter{equation}{0}
\renewcommand{\thesection}{\Alph{section}}
\section*{Appendix \thesection\protect\indent \parbox[t]{11.15cm}{#1}}
\addcontentsline{toc}{section}{Appendix \thesection\ \ \ #1}}

\def\bbe{{\bf{e}}}

\font\mybb=msbm10 at 11pt

\def\bb#1{\hbox{\mybb#1}}

\def\bR {\bb{R}}

\newcommand{\tI}{\text{\tiny $I$}}

\newcommand{\tA}{\text{\tiny $A$}}

\newcommand{\bea}{\begin{eqnarray}}
\newcommand{\eea}{\end{eqnarray}}

\usepackage[usenames,dvipsnames,svgnames,table]{xcolor}	% AH -- REMOVE BEFORE UPLOAD
\usepackage[normalem]{ulem}				% AH -- REMOVE BEFORE UPLOAD
\usepackage{microtype}
\usepackage[colorlinks, citecolor=blue, linkcolor=blue, urlcolor=Maroon, filecolor=Maroon, linktocpage=true]{hyperref} % AH

\usepackage{chngcntr}
\counterwithout{equation}{section}

\begin{document}

\begin{center}
%\today
\vspace*{-1.0cm}
\begin{flushright}
%\normalsize{\texttt{ZMP-HH/17-24}}\\
\end{flushright}
%\hfill hep-th/yymmnnn \\
%\hfill UB-ECM-PF-06-43 \\
%\hfill DMUS--MP--13/06 \\

\vspace{2.0cm} {\Large \bf A uniqueness theorem for warped $N>16$ Minkowski backgrounds with fluxes} \\[.2cm]

\vskip 2cm
 S.  Lautz$^1$ and  G.  Papadopoulos$^{1,2}$\footnotetext{GP is on study leave from the Department of Mathematics, King's College London.}
\\
\vskip .6cm

\begin{small}
${}^1$ \textit{Department of Mathematics, King's College London
\\
Strand, London WC2R 2LS, UK}\\
\texttt{sebastian.lautz@kcl.ac.uk}\\
\end{small}
\vskip0.5cm

\begin{small}
${}^2$\textit{Theoretical Physics Department
\\
CERN
\\
1211 Geneva 23, Switzerland}\\
\texttt{george.papadopoulos@kcl.ac.uk}
\end{small}
\\*[.6cm]

\end{center}

\vskip 2.5 cm

\begin{abstract}
\noindent
We demonstrate that  warped Minkowski space backgrounds, $\bR^{n-1,1}\times_w M^{d-n}$, $n\geq3$,   that preserve strictly more than 16 supersymmetries in $d=11$ and type II $d=10$ supergravities and with fields which may not be smooth everywhere are locally isometric to the $\bR^{d-1,1}$ Minkowski vacuum.   In particular, all such flux compactification  vacua of  these theories  have the same local geometry as the  maximally supersymmetric vacuum $\bR^{n-1,1}\times T^{d-n}$.

\end{abstract}

\newpage

%\tableofcontents

%%%%%%%%%%%%%%%%%%%%%%%%%%%%%%%%%%%%%%%%%%%%%%%%%%%%%%%%%%%%%%%%%%%%%%%%%%
%\setcounter{section}{0}\setcounter{equation}{0}
%%%%%%%%%%%%%%%%%%%%%%%%%%%%%%%%%%%%%%%%%%%%%%%%%%%%%%%%%%%%%%%%%%%%%%%%%%
%\section{Introduction}

Recently,  all  warped anti-de-Sitter  (AdS) backgrounds with fluxes that preserve $N>16$ supersymmetries in $d=11$ and $d=10$ supergravities  have been classified up to a local isometry
in \cite{sbjggp, ahslgp1, ahslgp2}.  In this note, we extend this result to include all warped $\bR^{n-1,1}\times_w M^{d-n}$   backgrounds of these theories.  In particular, we demonstrate that all warped $\bR^{n-1,1}\times_w M^{d-n}$, $n\geq3$, solutions with fluxes of $d=11$,  IIA $d=10$  and IIB $d=10$  supergravities that preserve $N>16$ supersymmetries are locally isometric to the $\bR^{d-1,1}$ maximally supersymmetric vacuum of these theories. Massive IIA supergravity does not admit such solutions.   A consequence of this
is that all  $N>16$ flux compactification vacua, $\bR^{n-1,1}\times_w M^{d-n}$,  of these theories   are locally isometric to the maximally supersymmetric toroidal vacuum $\bR^{n-1,1}\times T^{d-n}$. To prove these results we have made an  assumption that  the translation isometries along the  $\bR^{n-1,1}$ subspace of these backgrounds commute with all the odd generators of their Killing superalgebra.  The necessity and justification of this assumption will be made clear below.

To begin, we shall first describe the steps of the proof of our result which are common   to  all $d=11$ and $d=10$ theories and then specialize at the end to present the special features of the proof for each individual theory.  Schematically, the fields of $\bR^{n-1,1}\times_w M^{d-n}$ backgrounds are
\bea
ds^2&=&A^2 ds^2(\bR^{n-1,1})+ds^2 (M^{d-n})~,
\cr
F&=& W\wedge \mathrm{dvol}_A(\bR^{n-1,1})+ Z~,
 \eea
 where $A$ is the warp factor that depends only on the coordinates of the internal space $M^{d-n}$, $\mathrm{dvol}_A(\bR^{n-1,1})$ denotes the volume form of $\bR^{n-1,1}$ evaluated in the warped metric and $F$ denotes collectively all the k-form fluxes of the
 supergravity theories. We take $ds^2(\bR^{n-1,1})=2dudv+dz^2+\delta_{ab} dx^a dx^b$, where we have singled out a spatial coordinate $z$ which will be useful later. $W$ and $Z$ are $(k-n)-$ and $k-$ forms on $M^{d-n}$ which depend only on the coordinates of $M^{d-n}$.    Clearly if $n>k$,  $W=0$. Therefore these backgrounds are invariant under the Poincar\'e isometries of the $\bR^{n-1,1}$ subspace. It is known that there are no  smooth compactifications with non-trivial fluxes  of $d=10$ and $d=11$ supergravities \cite{gibbonsa, nunez}, i.e. solutions for which all fields are smooth including the warp factor and $M^{d-n}$ is compact without boundary.  However here we do not make these assumptions. $M^{d-k}$ is allowed to be non-compact and the fields may not be smooth.

 To continue following  the description of $\bR^{n-1,1}\times_w M^{d-n}$ backgrounds in \cite{sbjggp1,sbjggp2, sbjggp3}, where one can also find more details about our notation, we  introduce a light-cone orthonormal frame
 \bea
 \bbe^+= du~,~~~ \bbe^-= (dr-2 r A^{-1} dA)~,~~~ \bbe^m=A\, dx^m~,~~~ \bbe^i=e^i_\tI dy^\tI~,
 \label{orthonorm}
  \eea
 on the spacetime with $ds^2(M^{d-n})=\delta_{ij} \bbe^i \bbe^j$.   Then the Killing spinors of the $\bR^{n-1,1}\times_w M^{d-n}$ backgrounds can be written  as
\begin{eqnarray}
\epsilon&=&\sigma_++ u \Gamma_+\Gamma_z\Xi^{(-)}\sigma_-+A\sum_m x^m\Gamma_m  \Gamma_z\Xi^{(+)}\sigma_+
\cr
&&~~+\sigma_-+r \Gamma_-\Gamma_z\Xi^{(+)} \sigma_++ A\sum_m x^m\Gamma_m  \Gamma_z\Xi^{(-)}\sigma_-~,
\label{minksp1}
\end{eqnarray}
where $x^m=(z, x^a)$,  all the gamma matrices are in the frame basis (\ref{orthonorm}) and the spinors $\sigma_\pm$, $\Gamma_\pm\sigma_\pm=0$, depend only on the coordinates of $M^{d-n}$.  The remaining independent  KSEs are
a restriction of the gravitino and algebraic KSEs of the supergravity theories on $\sigma_\pm$ which schematically can be written as
\bea
D^{(\pm)}_i\sigma_\pm=0~,~~~{\cal A}^{(\pm)}\sigma_\pm=0~,
\label{rkse}
\eea
respectively,
and in addition an integrability condition
\bea
(\Xi^{(\pm)})^2\sigma_\pm=0~,
\label{minkremkses}
\eea
where $\Xi^{(\pm)}$ is a Clifford algebra element that depends on the fields. $\Xi^{(\pm)}$ will not be given here and can be found in the references above. The latter arises as a consequence of integrating the gravitino KSE of the theories along the $\bR^{n-1,1}$ subspace.

Notice that the (spacetime) Killing spinors (\ref{minksp1}) may depend on the coordinates of the $\bR^{n-1,1}$ subspace. Such a dependence  arises whenever $\sigma_\pm$ is not in the kernel of $\Xi^{(\pm)}$.  Of course $\sigma_\pm$ is required to lie in the kernel of
of $(\Xi^{(\pm)})^2$.  To see why this dependence can arise for $\bR^{n-1,1}\times_w M^{d-n}$ backgrounds, notice that $AdS_{n+1}$ in Poincar\'e coordinates
can be written as a warped product of $\bR^{n-1,1}\times_w \bR$.  Therefore all  AdS backgrounds, warped or otherwise, can be interpreted as warped
Minkowski space backgrounds.  It is also known that the former admit Killing spinors that depend on all AdS coordinates including those of
the Minkowski subspace. Therefore $\bR^{n-1,1}\times_w M^{d-n}$ may also admit  Killing spinors that depend on the coordinates of $\bR^{n-1,1}$, see also \cite{ugjggp4} for a more detailed explanation.

The assumption we have made that the commutator  of the translations $P$ along $\bR^{n-1,1}$ and the odd generators $Q$ of the
Killing superalgebra \cite{pkt, jose1} must vanish, $[P,Q]=0$, is required for  Killing spinors $\epsilon$  not to exhibit a dependence on the coordinates of $\bR^{n-1,1}$. Indeed, if the Killing spinors have a dependence on the Minkowski subspace coordinates, then the commutator $[P,Q]$ of the Killing superalgebra  will not vanish. This can be verified with an explicit computation of the spinorial Lie
derivative of $\epsilon$ in (\ref{minksp1})  along the translations of $\bR^{n-1,1}$.   Although this may seem as a technical assumption, it also has a physical significance in the context of flux compactifications.  Typically the reduced theory is invariant under the Killing superalgebra of the  compactification vacuum. So for the reduced theory to exhibit at most super-Poincar\'e invariance,  one must set $[P,Q]=0$ for all $P$ and $Q$ generators.
 This physical justification applies only to compactification vacua but we shall take it to be valid for all backgrounds that we shall investigate below. Of course such an assumption excludes all AdS solutions of supergravity theories re-interpreted as  warped Minkowski backgrounds.  Therefore from now on we shall take $\Xi^{(\pm)}\sigma_\pm=0$ and so {\it all} Killing spinors $\epsilon$ will not depend on
the coordinates of the $\bR^{n-1,1}$ subspace.

Before we proceed further, let us describe the Killing spinors of $\bR^{n-1,1}\times_w M^{d-n}$  backgrounds in more detail.  It turns out that if $\sigma_+$ is a Killing spinor, then $\sigma_-\defeq A\Gamma_{-z} \sigma_+$ is also a Killing spinor.  Similarly if $\sigma_-$ is a Killing spinor,  then $\sigma_+ \defeq A^{-1} \Gamma_{+z} \sigma_-$ is  also a Killing spinor.  Furthermore if $\sigma_+$ is a Killing spinor, then  $\sigma'_+\defeq\Gamma_{mn} \sigma_+$ are also Killing spinors for every $m,n$. Therefore the Killing spinors form multiplets under these Clifford algebra operations.  The counting of Killing spinors of a background  proceeds with identifying the linearly  independent Killing spinors in each multiplet and then counting the number of different multiplets that can occur \cite{sbjggp1,sbjggp2, sbjggp3}. As all Killing spinors are generated from $\sigma_+$ Killing spinors, we shall express all key formulae in terms of the latter.

The 1-form bilinears that are associated with spacetime Killing vectors $\epsilon^r$, $r=1,\dots, N$,  that  also leave all other fields invariant are
\bea
X(\epsilon^r, \epsilon^s)\defeq \langle (\Gamma_+-\Gamma_-) \epsilon^r, \Gamma_\tA \epsilon^s\rangle_s\, \bbe^\tA~,
\eea
where in $d=11$ and IIA supergravity theories  $\langle(\Gamma_+-\Gamma_-)\cdot, \cdot\rangle_s$ is the Dirac inner product restricted on the Majorana representation of $Spin(10,1)$ and $Spin(9,1)$, respectively, while in IIB it is the real part of the Dirac inner product.
Note that $X(\epsilon^r, \epsilon^s)=X(\epsilon^s, \epsilon^r)$.
In particular, one finds that
\bea
X(\sigma_-^r, \sigma_-^s)&=& 2 A^2\langle \sigma_+^r, \Gamma_z\Gamma_\tA \Gamma_-\Gamma_z\sigma_+^s\rangle_s \bbe^\tA~,
\cr
X(\sigma^r_-, \sigma_+^s) &=&2 A \langle \sigma_+^r, \Gamma_z\Gamma_\tA \sigma_+^s\rangle_s \bbe^\tA~,
\cr
X(\sigma_+^r, \sigma_+^s)&=& -\langle \sigma_+^r, \Gamma_+\Gamma_\tA \sigma_+^s\rangle_s \bbe^\tA~.
\label{bilin}
\eea
Clearly the last 1-form bilinear  is $X(\sigma_+^r, \sigma_+^s)= -2 \langle \sigma_+^r, \sigma_+^s\rangle_s \bbe^-$.  The requirement that $X(\sigma_+^r, \sigma_+^s)$  is Killing  implies that $\langle \sigma_+^r, \sigma_+^s\rangle_s$ are constants. In particular, one can choose without loss of generality that $\langle \sigma_+^r, \sigma_+^s\rangle_s=(1/2) \delta^{rs}$.  Then first 1-form bilinear in (\ref{bilin}) is $X(\sigma_-^r, \sigma_-^s)=2 A^2 \delta^{rs} \bbe^+$.

Next consider the middle 1-form bilinear in (\ref{bilin}). If $\sigma_+^r$ is in the same multiplet as $\sigma_+^s$, i.e. $\sigma_+^s=\Gamma_{za} \sigma_+^r$, then $X(\sigma^r_-, \sigma_+^s)=-\delta^{rs} A\, \bbe^a$. On the other hand if   $\sigma_+^s=\sigma_+^r$, then
$X(\sigma^r_-, \sigma_+^r)=A\, \bbe^z$.  Thus all these bilinears generate the translations in $\bR^{n-1,1}$.  However, if $\sigma_+^r$ and $\sigma_+^s$ are not in the same multiplet, then the bilinear
\bea
\tilde X_{rs}\defeq X(\sigma^r_-, \sigma_+^s) =2 A \langle \sigma_+^r, \Gamma_z\Gamma_i \sigma_+^s\rangle_s\, \bbe^i~,
\label{inkill}
\eea
will generate the isometries of the internal space.  The Killing condition of $\tilde X$ implies that
\bea
\tilde X_{rs}^i\partial_i A=0~.
\label{txia}
\eea
As the $\tilde X$ isometries commute with the translations,  the even part, $\mathfrak{g}_0$, of the Killing superalgebra decomposes as $\mathfrak{g}_0=\mathfrak{p}_0\oplus \mathfrak{t}_0$, where $\mathfrak{p}_0$ is the Lie algebra of translations in $\bR^{n-1,1}$ and $\mathfrak{t}_0$ is the Lie algebra of isometries in the internal space $M^{d-n}$.

So far we have not used the assumption that the backgrounds preserve $N>16$ supersymmetries. If this is the case, the Killing vectors generated by $\mathfrak{g}_0$ span the tangent space of the spacetime at each point.  This is a consequence of the homogeneity theorem proven for $d=11$ and $d=10$ supergravity backgrounds in \cite{jfofh1, jfofh2}.  This states that all solutions of these theories that preserve more than 16 supersymmetries must be locally homogeneous. In this particular case because of the decomposition of $\mathfrak{g}_0$, the Killing vector fields generated by $\mathfrak{t}_0$ span the tangent space of $M^{d-n}$ at every point. As a result   the condition (\ref{txia}) implies that $A$ is constant.    The main result of this note then follows as a consequence of the field equation of the warp factor and those of the rest of the scalar fields of these theories.  So to complete the proof we shall state the relevant equations on a case by case basis.

In $d=11$ supergravity, the 4-form field strength of the theory for   $\bR^{n-1,1}\times_w M^{11-n}$, $n\geq 3$, backgrounds can be expressed as
\bea
F= d\mathrm{vol}_A(\bR^{n-1,1})\wedge W^{4-n}+ Z~,
\eea
with $W^{4-n}=0$ for $n>4$, and the warp factor field equation is
\bea
\tilde \nabla^2 \log A=-n (\partial\log A)^2+{1\over 3\cdot (4-n)!} (W^{4-n})^2+{1\over 144} Z^2~,
\eea
where $\tilde \nabla$ is the Levi-Civita connection on $M^{11-n}$. The superscripts on the forms denote their degree whenever it is required for clarity. Clearly if $A$ is constant, as it has been demonstrated above for $N>16$ backgrounds, then $W^{n-4}=Z=0$.  So $F=0$ and thus all the fluxes vanish.  In fact, this is also the case for $n=2$ provided that $A$ is taken to be constant.
As $F=0$ and $A$ is constant, the gravitino KSE  in (\ref{rkse}) implies that all the Killing spinors $\sigma_\pm$ are  parallel with respect to the Levi-Civita connection, $\tilde \nabla$, on $M^{11-n}$.  In turn this gives that all the Killing vector fields $\tilde X$ in (\ref{inkill}),  which span the tangent space of $M^{11-n}$,  are also parallel with respect to $\tilde \nabla$.  Thus  $M^{11-n}$  is locally isometric to  $\bR^{11-n}$.  Therefore the backgrounds
$\bR^{n-1,1}\times_w M^{11-n}$ are locally isometric to the maximally supersymmetric vacuum $\bR^{10,1}$. Notice that the last step of the proof requires the use of
the homogeneity theorem.

In (massive) IIA supergravity, the   4-form $F$, 3-form $H$ and  2-form $G$ field strengths of the theory for  $\bR^{n-1,1}\times_w M^{10-n}$, $n\geq 3$, backgrounds  can be written as
\bea
F&=&\mathrm{dvol}_A(\bR^{n-1,1})\wedge W^{4-n}+ Z~,
\cr
H&=&\mathrm{dvol}_A(\bR^{n-1,1})\wedge P^{3-n}+ Q~,
\cr
G&=& L~,
\eea
where $W^{4-n}$ vanishes for $n>4$ and similarly $P^{3-n}$ vanishes for $n>3$. The
 field equations for  the warp factor $A$ and dilaton field $\Phi$, $n>2$, are
\bea
\tilde\nabla^2\log\, A&=&- n (\partial\log\, A)^2+2 \partial_i\log\, A \partial^i\Phi+{1\over 2} (P^{3-n})^2+{1\over 4} S^2+{1\over 8} L^2+{1\over 96} Z^2+{1\over 4} (W^{4-n})^2~,
\cr
\tilde\nabla^2\Phi&=&-n \partial_i\log A \partial^i\Phi+ 2 (d\Phi)^2-{1\over12} Q^2+{1\over2} (P^{3-n})^2+{5\over 4} S^2+{3\over 8} L^2+{1\over 96} Z^2
\cr && \qquad-{1\over 4} (W^{4-n})^2~,
\eea
where $S=e^\Phi m$ and $m$ is the cosmological constant of (massive) IIA supergravity.
Clearly if both $A$ and $\Phi$ are constant, which is the case for all $N>16$ backgrounds,  then the above two field equations imply that all the form fluxes will vanish.  Significantly, the cosmological constant must vanish as well. There are no $\bR^{n-1,1}\times_w M^{10-n}$ solutions in massive IIA supergravity that preserve $N>16$ supersymmetries.  In IIA supergravity, an argument similar to the one presented above in $d=11$ supergravity reveals
that $M^{10-n}$ is locally isometric to $\bR^{10-n}$ and so all $N>16$  $\bR^{n-1,1}\times_w M^{10-n}$ backgrounds are locally isometric to the maximally
supersymmetric vacuum $\bR^{10,1}$.   It is not apparent that the theorem holds for $n=2$ even if $A$ and $\Phi$ are taken to be constant.

In IIB supergravity  the self-dual real  5-form $F$ and complex  3-form $H$  field strengths of the theory for  $\bR^{n-1,1}\times_w M^{10-n}$, $n\geq 3$, backgrounds  can be expressed as
\bea
F&=&\mathrm{dvol}_A(\bR^{n-1,1})\wedge W^{5-n}+ \star W^{5-n}~,
\cr
H&=&\mathrm{dvol}_A(\bR^{n-1,1})\wedge P^{3-n}+ Q~,
\eea
where $W^{5-n}$ vanishes for $n>5$ and $P^{3-n}$ vanishes for $n>3$.
The field equation of the warp factor is
\bea
\tilde\nabla^2\log A=-n (\partial\log A)^2+{3\over 8} |P^{3-n}|^2+{1\over 48} |Q|^2+{4\over (5-n)!} (W^{5-n})^2~.
\eea
Clearly if $A$ is constant, which we have demonstrated that it holds for $N>16$ backgrounds, then $F=H=0$.  Moreover the homogeneity theorem implies that the two scalar fields of
IIB supergravity, the axion and the dilaton, are also constant. A similar argument to that used in $d=11$ supergravity
implies  that $M^{10-n}$ is locally isometric to $\bR^{10-n}$.  Thus $\bR^{n-1,1}\times_w M^{10-n}$ are locally isometric
to the $\bR^{9,1}$ maximally supersymmetric vacuum of the theory. The same conclusion holds for $n=2$ as well provided that $A$ is taken to be constant.

%\section{Conclusions}
%%%%%%%%%%%%%%%%%%%%%%%%%%%%%%%%%%%%%%%%%%%%%%%%%%%%%%%%%%%%%%%%%%%%%%%%%%%%%%%%%%%%%%%%%%%%%%%%
\section*{Acknowledgments}

 GP would like to thank the Theoretical Physics Department at CERN for hospitality and support. GP is partially supported from the  STFC rolling grant ST/J002798/1.

%%%%%%%%%%%%%%%%%%%%%%%%%%%%%%%%%%%%%%%%%%%%%%%%%%%%%%%%%%%%%%%%%%%%%%%%%%%%%%%%%%%%%%%%%%%%%%%%
%\setcounter{section}{0}\setcounter{equation}{0}

%\appendix{Notation and conventions}

\end{document}